\documentclass[11pt]{article}

\usepackage[top=2.8cm,left=2.5cm,bottom=3.2cm,right=2.5cm]{geometry}

\usepackage[english]{babel}
\usepackage{amsmath, amssymb}

\usepackage{cite}
\usepackage{authblk}
\usepackage[font={small}]{caption}

\usepackage{microtype}

\usepackage{tikz}
\usetikzlibrary{matrix,calc,intersections,decorations.pathreplacing,positioning}

\usepackage[english]{babel}

\newenvironment{definition}[1][Definition]{\begin{trivlist}
\item[\hskip \labelsep {\bfseries #1}]}{\end{trivlist}}

\begin{document}


\title{Measuring logic complexity can guide pattern discovery\\ in empirical systems}

\author[1,2]{Marco Gherardi\footnote{marco.gherardi@mi.infn.it}}
\author[1,2]{Pietro Rotondo}
\affil[1]{Universit\`a degli Studi di Milano, via Celoria 16, 20133 Milano, Italy}
\affil[2]{Istituto Nazionale di Fisica Nucleare, sezione di Milano, Italy}

\maketitle

\begin{abstract}
We explore a definition of complexity based on logic functions,
which are widely used as compact descriptions of rules
in diverse fields of contemporary science.
Detailed numerical analysis shows that
(i) logic complexity is effective in discriminating between classes of functions
commonly employed in modelling contexts;
(ii) it extends the notion of canalisation, 
used in the study of genetic regulation,
to a more general and detailed measure;
(iii) it is tightly linked to the resilience of a function's output 
to noise affecting its inputs.
We demonstrate its utility
by measuring it in empirical data on gene regulation,
digital circuitry, and propositional calculus.
Logic complexity is exceptionally low in these systems.
The asymmetry between ``on'' and ``off'' states
in the data correlates with the complexity in a non-null way;
a model of random Boolean networks clarifies this trend
and indicates a common hierarchical architecture
in the three systems.
\end{abstract}


\vspace{1cm}


\section{Introduction}

Irreducibility is a property often ascribed to complex
entities:
their behaviour can not be compressed into compact descriptions.
Symmetries --- a paramount concept in physics ---
and recurrent patterns are prominent facilitators,
enabling implicit definitions of the systems.
The amount of implicitness allowed is a measure of
information content \cite{Carbone:BOOK}, 
as illustrated by Kolmogorov's definition of complexity.
The concept of complexity pervades contemporary science,
from the statistical mechanics of disordered systems and complex networks
to economical and technological studies,
prominently in the emerging field of complex systems,
where it can promote the discovery of
regularities and large-scale trends
\cite{SteinNewman:BOOK,
BarratBarthelemyVespignani:BOOK,
HidalgoHausmann:2009,
TacchellaCristelli:2012,
TononiEdelman:1998,
McNerneyFarmer:2011,
Frenken:2006,
FeldmanCrutchfield:1998,
ShaliziShalizi:2004}.
A recurrent question,
notably in evolutionary biology,
is whether complexity contributes to fitness
\cite{AuerbachBongard:2014,JoshiTononi:2013,WangLiao:2010,AdamiOfria:2000}.
Its deep interplay with system-level properties such
as tolerance and modularity has been investigated
both in technological and biological designs
\cite{CseteDoyle:2002,CarlsonDoyle:2002};
robustness of complex systems against fluctuations and attacks, in particular,
is the subject of numerous studies
\cite{AlbertJeong:2000,MaciaSole:2009,LiLong:2004}.
Although complexity has been examined in detail
in single contexts,
by employing specific definitions,
a study across disciplines is still lacking,
and the general consequences and relations
with other traits are still largely unknown.
In this work we employ
a definition of complexity
based on Boolean logic
\cite{Wegener:BOOK},
that is generic enough to be applicable in diverse fields.
Logic functions are a natural and simple representation of how
information flows in complex systems
\cite{TkacikBialek:2014}.
They are used to express
genotype to phenotype mappings --- both in metabolic 
\cite{BarveWagner:2013}
and electronic
\cite{SoleValverde:2013}
systems ---,
rules for the control of gene expression \cite{HarrisSawhill:2002},
relations in protein networks \cite{Bowers:2004},
cryptographic cyphers
\cite{Carlet:2010},
functions realised by digital electronic circuits
\cite{Micheli:BOOK},
simple cooperative games
\cite{Bilbao:BOOK},
and they lie at the foundations of mathematical logic
\cite{Church:BOOK}.
Complexity in a Boolean setting has been addressed
especially regarding the global dynamics of Boolean networks
\cite{GongSocolar:2012},
yet it proves profitable already to focus on single functions (nodes)
\cite{CiandriniMaffi:2009,MarquesPitaMitchell:2008,MarquesPitaRocha:2011}.
We will concentrate on the properties of individual functions in this paper.

There are $2^{2^n}$ distinct
Boolean functions with $n$ variables:
finding useful coordinates in this high-dimensional space is a necessity
in all fields concerned with logic functions.
A large number of quantities describing various characteristics
have been defined and analysed.
Some are especially useful for assessing their cryptographic properties
(such as \emph{correlation immunity} \cite{Carlet:2010}),
some are suited for biological systems
(such as the \emph{canalising} quality \cite{HarrisSawhill:2002}),
some are designed to address issues in specific domains
(such as the \emph{Nakamura number} in cooperative game theory \cite{Nakamura:1979}).
Here we concentrate on two natural and general ``observables'',
bias and complexity,
whose versatility enables their use as a reference frame
for comparing different systems in different fields.

We define the notion of \emph{logic complexity}
of a Boolean function as the size of the most compact Boolean
expression that realises it \cite{Wegener:BOOK}
(see Sec.~\ref{section:definitions} for the choice of the description language).
Firstly (in Sec.~\ref{section:results}), 
we show that this definition assigns
quantitatively different complexities to
popular classes of logic functions.
We clarify its relation with bias ---
a measure of the asymmetry
between ``on'' and ``off'' values ---
and with resilience of the functions in noisy environments,
thus establishing a quantitative relation between
complexity and robustness.
Importantly, logic complexity realises a rigorous
and general measure for the notion of \emph{canalisation},
a fruitful concept developed in the context of gene regulation
\cite{HarrisSawhill:2002}.
In this field, our results further expose the inconveniences
of the commonly used threshold functions,
which turn out to have exceptionally high complexity.

Secondly (in Sec.~\ref{section:application}),
as an illustrative application of the concepts developed,
we compute the complexity and the bias
in three exemplary
systems belonging to the realms of biology,
technology, and mathematics,
namely genetic regulation, electronic circuits, and propositional calculus.
We find that the three systems are characterised by
different ranges of the bias,
and that the logic complexities are generally
small, compared to a null model of random Boolean functions.
The non-null trends are elucidated
by a model of random Boolean networks, suggesting
hierarchical organisation as a shared architecture.

Altogether, the results presented here advocate
the use of bias and complexity as coordinates in a ``morphospace''
for the classification of logic functions,
and in particular as a powerful tool for comparing
Boolean models and data.
More in general, our results remark that complexity
is a measurable and empirically relevant trait,
indicating similar features in dissimilar systems;
however, its role is entangled with other important properties,
such as bias, robustness, and information dispatching, 
and can not be contemplated in isolation.

\section{Measuring logic complexity and bias\label{section:definitions}}

\begin{figure}[t]
\centering
\begin{tikzpicture}[scale=1.]

\pgfmathsetmacro{\angl}{80}
\pgfmathsetmacro{\angr}{180-\angl}
\pgfmathsetmacro{\spc}{3em}
\pgfmathsetmacro{\dblspc}{4em}
\matrix (m) [matrix of math nodes,minimum width=1.5em,inner sep=0.4em] {
p&0&0&0&0&1&1&1&1\\
q&0&0&1&1&0&0&1&1\\
r&0&1&0&1&0&1&0&1\\
\hline
f(p,q,r)&1&1&0&1&1&1&0&1\\
};

\node[circle,draw,thick,minimum size=1.4em,color=blue] (no1) at (m-4-2) {};
\node[circle,draw,thick,minimum size=1.4em,color=blue] (no2) at (m-4-3) {};
\node[circle,draw,thick,minimum size=1.4em,color=red] (no3) at (m-4-5) {};
\node[circle,draw,thick,minimum size=1.4em,color=blue] (no4) at (m-4-6) {};
\node[circle,draw,thick,minimum size=1.4em,color=blue] (no5) at (m-4-7) {};
\node[circle,draw,thick,minimum size=1.4em,color=red] (no6) at (m-4-9) {};
 
\draw[color=red,thick] (no3) to[out=-\angl,in=-\angr,distance=\dblspc] (no6);

\coordinate (ey) at (0,1em);
\coordinate (ex) at (1em,0);
\path[name path=li1] ($(no2)-4*(ey)$) to (no2);
\path[name path=li2] ($(no4)-4*(ey)$) to (no4);

\path[name path=licur,draw,color=blue,thick] (no1) to[out=-\angl,in=-\angr,distance=\spc] (no5);
\path [name intersections={of = li1 and licur}]; 
\coordinate (i1) at (intersection-1) {};
\path [draw,color=blue,thick] (i1) to (no2);

\path [name intersections={of = li2 and licur}];
\coordinate (i2) at (intersection-1) {};
\path [draw,color=blue,thick] (i2) to (no4);

\node (f0) at ($0.5*(m-4-4)+0.5*(m-4-5)-2.*(ex)-5.7*(ey)$) {$=$};
\node [color=blue] (f1) at ($0.5*(m-4-4)+0.5*(m-4-5)-0.3*(ex)-5.7*(ey)$) {$(\neg q)$};
\node (f2) at ($(m-4-7)-5.7*(ey)-2.25*(ex)$) {$\vee$};
\node [color=red] (f3) at ($(m-4-7)-5.7*(ey)$) {$(q \wedge r)$};

\path [decoration={brace},decorate,draw] ($(m-1-9)+1*(ex)$) to ($(m-3-9)+1*(ex)$);
\path [decoration={brace},decorate,draw] ($(m-1-2)+1*(ey)$) to ($(m-1-9)+1*(ey)$);
\path [decoration={brace},decorate,draw] ($(m-4-9.north)+1*(ex)$) to ($(m-4-9.south)+1*(ex)$);
\path [decoration={brace},decorate,draw] ($(f3.north)+2*(ex)$) to ($(f3.south)+2*(ex)$);

\node[above = 1.1em of m-1-5.east] (n) {$2^n=8$};
\node[right = 0.4em of m-2-9] (n) {$n=3$};
\node[right = 0.4em of m-4-9] (B) {$B=\frac{6}{8}$};
\node[right = 0.4em of f3] (C) {$C=\frac{2}{8}$};

\path [->,draw] (m-4-1) to[out=-90,in=180] (f0);
\path [->,thick,color=blue,draw] (f1) to ($(f1)+2.5*(ey)$);
\path [->,thick,color=red,draw] (f3) to ($(f3)+1.8*(ey)$);

\end{tikzpicture}
\caption{\label{figure:boolean}
A Boolean function $f$ of $n=3$ literals $p,q,r$
has $8$ possible input combinations; the value
of $f(p,q,r)$ on each of these (its truth table)
completely specifies the function.
The fraction of combinations on which $f$ is true,
in this case 6 out of 8, is the bias $B$.
The full disjunctive normal form of $f$ is obtained
by explicitly stating all these combinations;
in this example it would be
$(\neg p \wedge \neg q \wedge \neg r) \vee (\neg p \wedge \neg q \wedge r)
\vee (\neg p \wedge q \wedge r) \vee (p \wedge \neg q \wedge \neg r)
\vee (p \wedge \neg q \wedge r) \vee (p \wedge q \wedge r)$.
However a shorter form can be obtained by dividing the support of $f$
into \emph{cells}, thus exploiting its symmetries.
In this case $2$ cells (blue and red groupings) are sufficient (and necessary),
thus the complexity is $C=2/8$.
}
\end{figure}
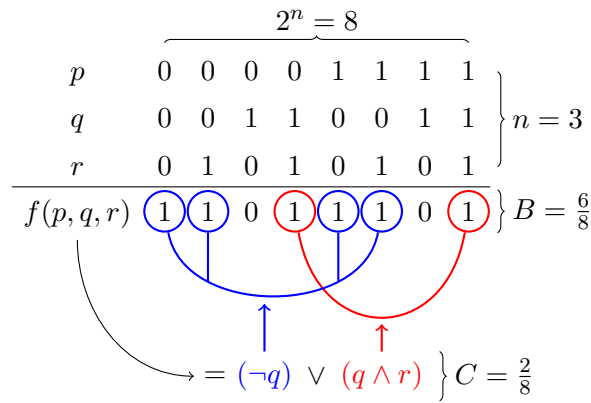

A \emph{Boolean} (or \emph{logic}) function maps the set $\{0,1\}^n$ to $\{0,1\}$,
associating a truth value to
each combination of its $n$ Boolean inputs
(by convention the integers 1 and 0 mean \emph{true} and \emph{false},
or \emph{on} and \emph{off}, respectively).
The binary nature of this description is sometimes just an approximation
to a continuous or multi-valued empirical situation
(such as the expression levels of a gene),
but has the advantage of being simple to deal with.
Since the domain is finite, a function can be specified by exhaustively
listing the values it takes for all input combinations.
Such a list constitutes the \emph{truth table} of the function.
See Fig.~\ref{figure:boolean} for an example.

%
The bias $B$,
defined as the average output value over all input combinations,
measures the propensity of the system described by the given function
to be in one of the two states 0 and 1:
$B=1$ for a tautology (the function that is true for all values of its inputs),
$B=0$ for a falsity (the negation of a tautology).

Beyond the truth table, there is a way of writing
Boolean functions which makes them more intelligible
to humans, as opposed to computers.
In fact, one can assign names to the function's inputs ---
the \emph{literals} --- and decompose the function into
binary sub-functions (i.e., functions of two literals)
and negations (which are unary functions),
thus writing it in the form of a \emph{Boolean expression};
Fig.~\ref{figure:boolean} presents an example.
Such an expression contains the literals (possibly repeated),
parentheses, and binary and unary operators.
All Boolean functions can be expressed in this way,
as every Boolean function can be written as the composition
of binary and unary functions.
We will use the set of logical operators $\{\land,\lor,\lnot\}$
(i.e., AND, OR, and NOT) to express them.
As an example, consider the function of $3$ literals $p,q$, and $r$
that is true if one or two literals are true, and false otherwise.
With our choice of basic operators it can be written as
$(p\lor q\lor r)\land \lnot(p\land q\land r)$.

The definition of complexity that we shall employ for Boolean functions
uses \emph{disjunctive normal forms} (DNF) as the description language.
A formula is in DNF if it is of the form 
$(A\land B\land \cdots)\lor(C\land D\land \cdots)\lor\cdots$,
where $A,B,\ldots$ are literals or their negations.
For example, $(p\land q)\lor\lnot p$ is in disjunctive normal form,
while $\lnot(p\lor q)$ is not.
Consider the tautology of $n=2$ literals.
A particular normal form (called \emph{full} DNF) can be built 
by listing explicitly all these combinations,
thus obtaining the formula
$(p\land q)\lor(p\land \lnot q)\lor(\lnot p\land q)\lor(\lnot p\land \lnot q)$;
however, a more concise formula would be, for instance, $p\lor\lnot p$.
Such a level of conciseness is intuitively related to the
lack of complexity of the tautology. 
In general one expects
that the more complex a function is, the less compact it can be made.
We shall then define the complexity $C$ of a function as
the number of terms in the \emph{shortest} DNF specifying the function
(normalised by $2^n$).
This definition assigns minimum complexity to tautologies
and their negations.
At the other end of the spectrum, the \emph{parity function},
which counts the number of true literals modulo 2,
is \emph{balanced} --- meaning that $B=1/2$ ---
and has the largest possible complexity, namely $1/2$.
Note that in general $C\leq B$.
The main definitions and concepts are summarised in Fig.~\ref{figure:boolean}
for a simple function.
See the Appendix for detailed definitions.

\begin{figure}
\centering
\includegraphics[scale=1.1]{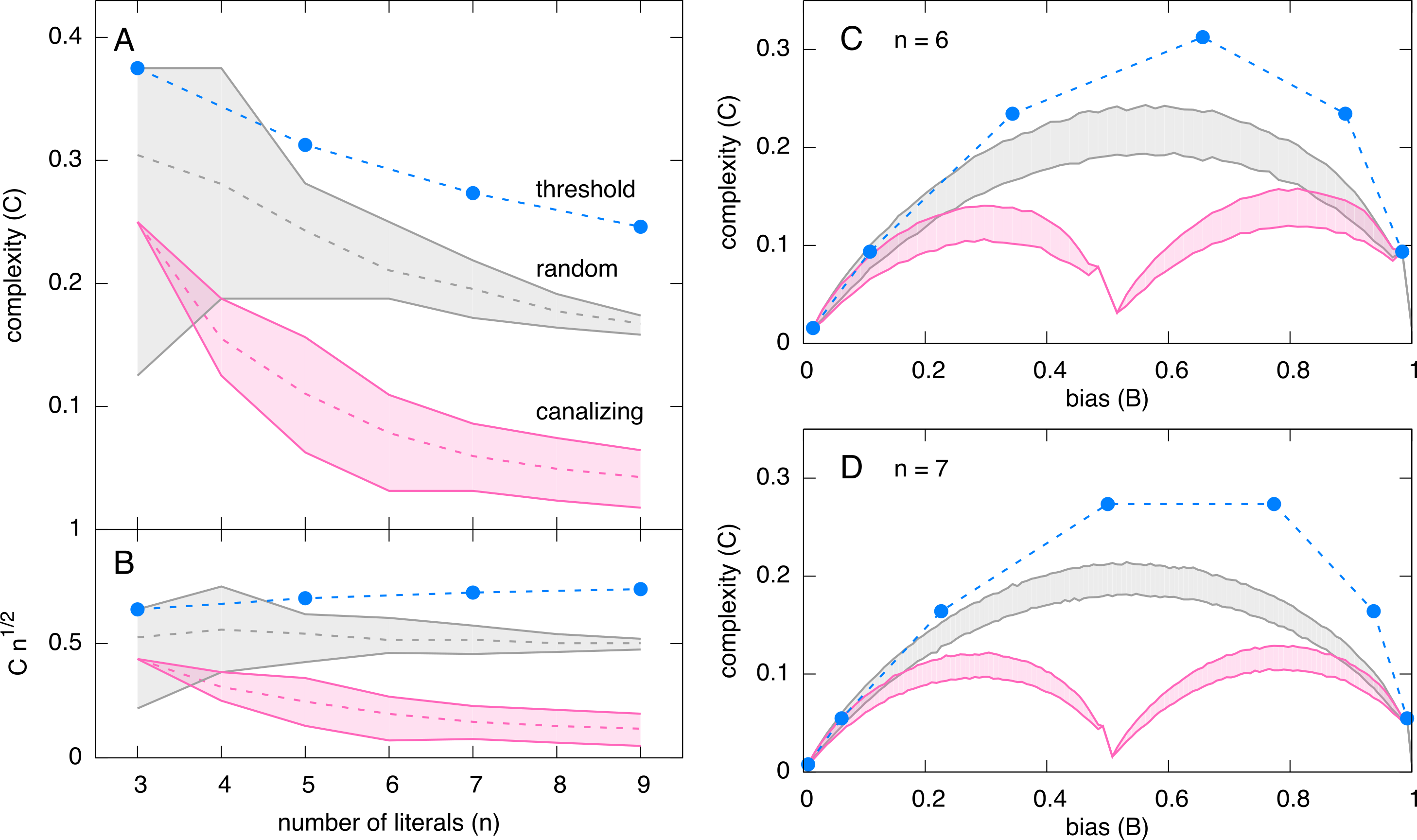}
\caption{\label{figure:classes}
Commonly-used classes of functions are characterized by markedly different
complexities.
Threshold functions are more complex than the typical functions,
while canalizing functions are simpler.
Shaded areas are the $80\%$ variability intervals
(at fixed bias) for each class.
(A)
Complexity as a function of the number of literals.
Bias is restricted to the interval $(0.4,0.6)$
(there are no threshold functions
of an even number of literals satisfying this constraint).
(B)
For random functions, $C$ decreases
approximately as $1/\sqrt{n}$. The plot shows
the rescaled quantity $C \sqrt{n}$.
(C,D)
Complexity as a function of bias for the three classes of functions considered,
with $n=6$ and $n=7$ literals.
}
\end{figure}

Measuring the bias of a function $f$ is straightforward,
as it amounts to counting the number of ones in the truth table.
Calculating the complexity, instead, is in general a
computationally hard problem.
The presence of symmetries in the truth table is what enables the compression.
By symmetry, in this context, we mean a choice of a particular combination of values
for a fixed subset of literals, such that the value of $f$,
conditioned to this choice, does not depend on the other literals.
This is the definition of a \emph{cell} (see the Appendix).
Finding the most compact normal form representing
a function (or \emph{minimising} it) thus
amounts to finding the smallest number of cells
sufficient to describe the function's truth values;
this means exploiting the set of its symmetries
in the best possible way.
However, even if one has listed all the cells
of a function, there could be non-trivial overlaps between them ---
causing what is known as \emph{frustration} in statistical mechanics ---
thus complicating the task of finding the smallest
subset that recovers the function.
In fact, this problem is equivalent to the ``set cover problem'',
a well-known NP-HARD problem in algorithmic complexity theory
\cite{Papadimitriou:2014,Korte:BOOK}.

Fortunately, since the minimisation of logic function is a crucial step in the
design of digital circuits,
standard algorithms are available for this task.
Here we use an implementation of the Quine-McCluskey method
\cite{Micheli:BOOK}, which
deterministically finds the minimal form of a function.
The maximum number of literals in the analyses presented here
is $n=9$.
For much larger functions, an approximate method is needed
(such as ``Espresso''
\cite{Espresso}).
Such methods have the advantage of being adapted to
multi-valued logic, and they even allow for indeterminacies, 
so they can be used for computing logic complexity in more
general settings.

We remark that ideas related to the one advanced here
were proposed in the fields of unconventional computation and cellular automata
\cite{MarquesPitaMitchell:2008,MarquesPitaRocha:2011},
where a notion of ``conceptual representation'' for Boolean rules was developed.
That method, which makes use of a representation in terms of cells,
essentially corresponds to the first step in the Quine-McCluskey algorithm,
before the set-covering problem is solved.

\section{Results\label{section:results}}

\subsection{Bias and logic complexity discriminate popular classes of functions}

We show here that the Boolean complexity
takes values lying in different ranges for different
commonly used classes of Boolean functions
\cite{Kauffman:BOOK}.
We examine \emph{random} functions,
i.e.~Boolean functions with a fixed number of literals
drawn with uniform probability,
\emph{canalising} functions,
for which the value of a single input variable
decides whether the other variables
have any influence on the result,
and \emph{threshold} functions,
for which the result is decided by the sum of
``enhancer'' variables minus the sum of
``inhibitor'' variables.

More formally, the ensemble of random functions with $n$ literals
is defined as the set of all $n$-variables Boolean functions,
endowed with the flat probability measure. This ensemble is useful
as a null (unconstrained) model for discerning positive features in other classes
of functions, as well as in the data.
Random canalising functions are defined as follows.
Consider a function $f(p_1,\ldots,p_n)$ of $n$ literals $p_1,\ldots,p_n$.
If $f$ is canalising, then by definition there exists a literal 
(by rearranging the literals, we can assume it is $p_1$)
and two truth values (the input $I$ and the output $O$) such that
$f(I,p_2,\ldots,p_n)=O$. If $p_1=\lnot I$, then
$f(\lnot I,p_2,\ldots,p_n)=g(p_2,\ldots,p_n)$,
where $g$ is a function of $n-1$ variables.
The random canalising ensemble is specified by taking
both $I$ and $O$ to be 0 or 1 independently with
the same probability, and $g$ to be a random (uniform) function of $n-1$ literals.
Threshold functions are
often used to model regulatory rules
starting from known molecular interactions
(e.g., in the cell cycle of yeast \cite{LiLong:2004,Davidich:2008}).
Let $a_j$ be a set of $n-1$ couplings, specifying the nature of
the influence of a protein, identified by $j$, on a given gene product, identified by $i$.
In particular, $a_j=1$ if $j$ is an enhancer and $a_j=-1$ if it is an inhibitor for $i$
(interaction strength can be taken into account by extending the possible
values of $a_j$).
If $x_j$ are on/off values specifying the presence ($1$) or absence ($0$)
of each protein,
then the Boolean state $f$ of $i$, is computed as $1$ or $0$
depending on the sign of $\sum_{j\neq i} a_j x_j-\theta$
(where $\theta$ is the threshold)
with the exception that if the value is zero then $f=x_i$.
The ensemble of random threshold functions of $n$ variables is defined
by taking the $a_j$'s to be random independent Bernoulli variables
in $\{-1,1\}$ (we will fix $\theta=0$, unless specified otherwise).

Figure~\ref{figure:classes}A shows the complexities of
the three classes of functions defined above,
as the number of inputs $n$ is varied.
For clarity, we excluded very unbalanced functions from this analysis,
by restricting biases to the interval $0.4<B<0.6$.
Threshold, canalizing, and random functions
segregate into separate regions, already for $n\gtrsim 5$.
Interestingly, the combination $C n^{1/2}$ appears to be
approximately increasing, decreasing, and constant in $n$
for the three classes respectively
(Fig~\ref{figure:classes}B).

Let us restrict the analysis to fixed numbers of literals $n$,
in order to explore the relations between bias and complexity.
Panels C and D in Fig.~\ref{figure:classes}
show how the three classes occupy different regions
in the B-C space.
These plots have been obtained by generating random
functions of each class for each possible value of $B$ fixed.
While all types of functions have comparable complexities
for extreme biases ($B\lesssim 0.2$ and $B\gtrsim 0.8$),
in the balanced regime they are sharply discriminated by complexity.
This rules out the possibility that
the trends observed above might be due solely to
how the typical biases $B$ depend on $n$
in the three classes of functions.
Notice that both the bias and the complexity of threshold functions are quantised,
because of symmetry under permutations of the variables.
Perhaps surprisingly, their complexity
is larger than the typical value.
Changing the threshold value $\theta$ has the only effect
of changing the ensemble weights of the functions
(larger $\theta$ favours functions with lower bias),
but the B-C plot remains the same.

The figure also exposes a non-monotonic correlation
between complexity and bias, common to random and threshold
functions.
Canalising functions, whose definition modularly uses
a random function of $n-1$ variables, present a similar
behaviour on a halved scale;
we are going to study this pattern more in detail.

\subsection{Logic complexity realises a quantitative measure of canalisation}

\begin{figure}
\centering
\includegraphics[scale=1.1]{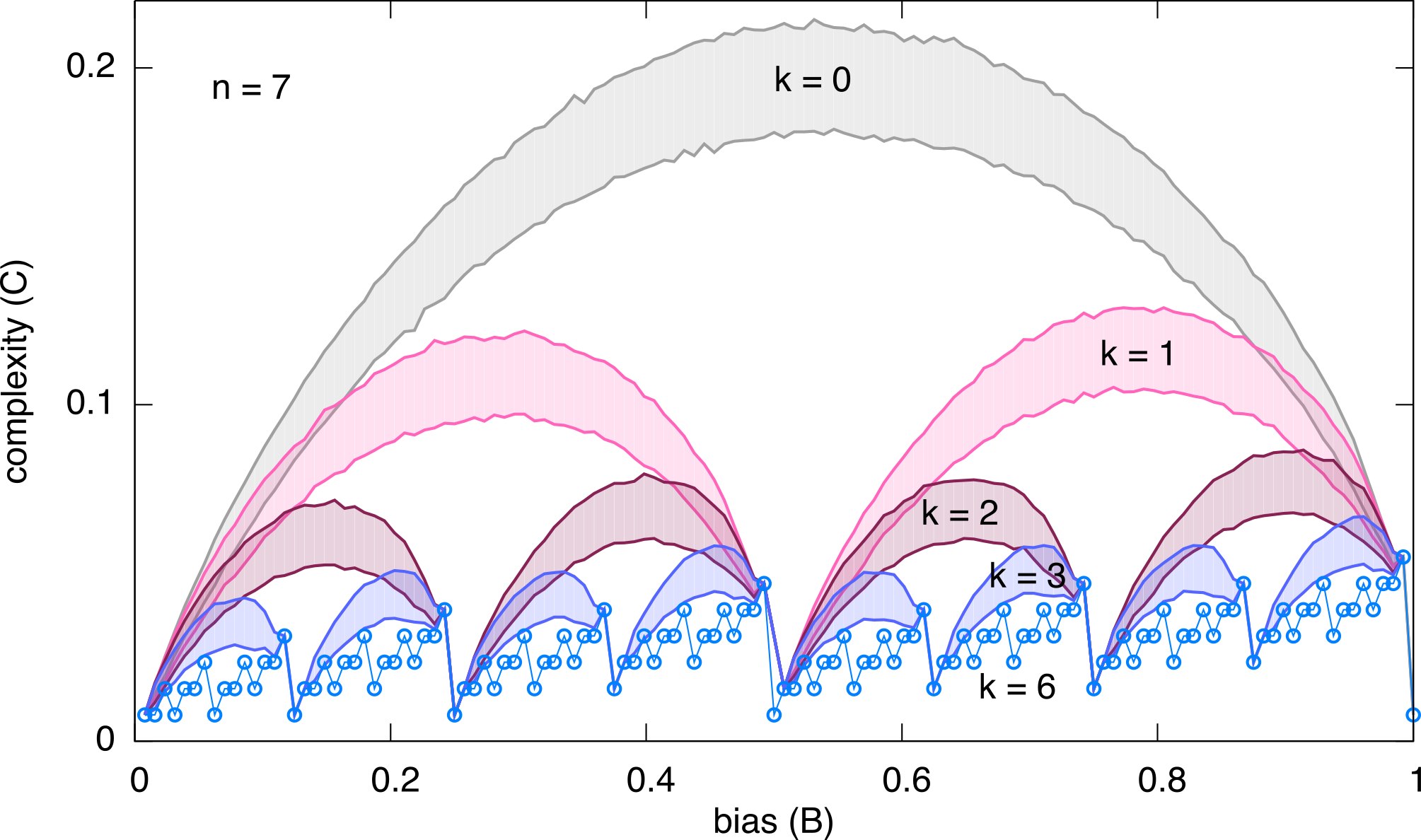}
\caption{\label{figure:canalizing}
The complexity of random nested canalising
functions decreases systematically
with the level $k$.
The shaded areas are the 80\%
variability intervals.
The inverted-U pattern repeats itself $2^k$
times at level $k$. 
At level $n-1$ (lowermost curve) complexity
is a deterministic function of bias
(circles are the analytical expression given in the text).
Here the number of input variables is $n=7$.
}
\end{figure}

The definition of canalisation employed above
isolates exactly one canalising variable $p_1$.
However, a more general definition can be given,
where the number of such special variables is $k<n$.
If the value of the first canalising variable does not fix
the output, then the second canalising variable is
considered, and so on in a nested fashion, until the $k$-th variable.
More precisely, given a set of input values $I_1,\ldots,I_k$
and a set of output values $O_1,\ldots,O_k$,
$f(p_1,\ldots,p_n)$ takes the value $O_j$ if $p_i\neq I_i$
for all $i<j$ and $p_j=I_j$;
otherwise it is a function $g(p_{k+1},\ldots,p_n)$ of the remaining $n-k$ variables.
The ensemble is specified by taking the $I_i$'s and $O_i$'s
as random independent Bernoulli variables in $\{0,1\}$
and $g$ as a random function of $n-k$ literals.
These functions will be called \emph{random nested canalising},
and $k$ their \emph{level}
(our definition is based on that of nested canalisation
given in \cite{KauffmanPeterson:2003}, but it uses a less constrained measure).

We generated 1000 random nested canalising functions 
of level $k=1,\ldots,n-1$ for each
possible value of their bias, and computed their complexity.
The results (with $n=7$) are in Fig.~\ref{figure:canalizing}
(not all levels are shown for clarity;
other values of $n$ yield similar results).
Disregarding for the moment the fine structure
that appears as a function of bias,
the overall trend is clear: the more levels of canalisation
a function has, the smaller is its complexity.
Canalisation itself can not be measured quantitatively:
the level $k$ is a rough measure, but it takes
only $n-1$ different values.
Therefore, complexity,
which can take $2^n$ different values,
appears as a much more detailed
measure of canalisation.

The fractal nature of the plot is interesting.
At level $k$, the inverted-U pattern displayed by random functions
is repeated $2^k$ times (this is true independently of $n$).
Fully canalising functions, i.e.~those at level $n-1$,
satisfy a deterministic relation between bias and complexity,
which can be seen to be given by
$C=S_2(B 2^n)/2^n$,
where $S_q(m)$ is the sum of all the
digits in the base-$q$ representation of the integer $m$
(in the case $q=2$ it is known as the binary weight).

\subsection{Logic complexity constrains robustness}
\label{section:robustness}

\begin{figure}[t]
\centering
\includegraphics[scale=1.1]{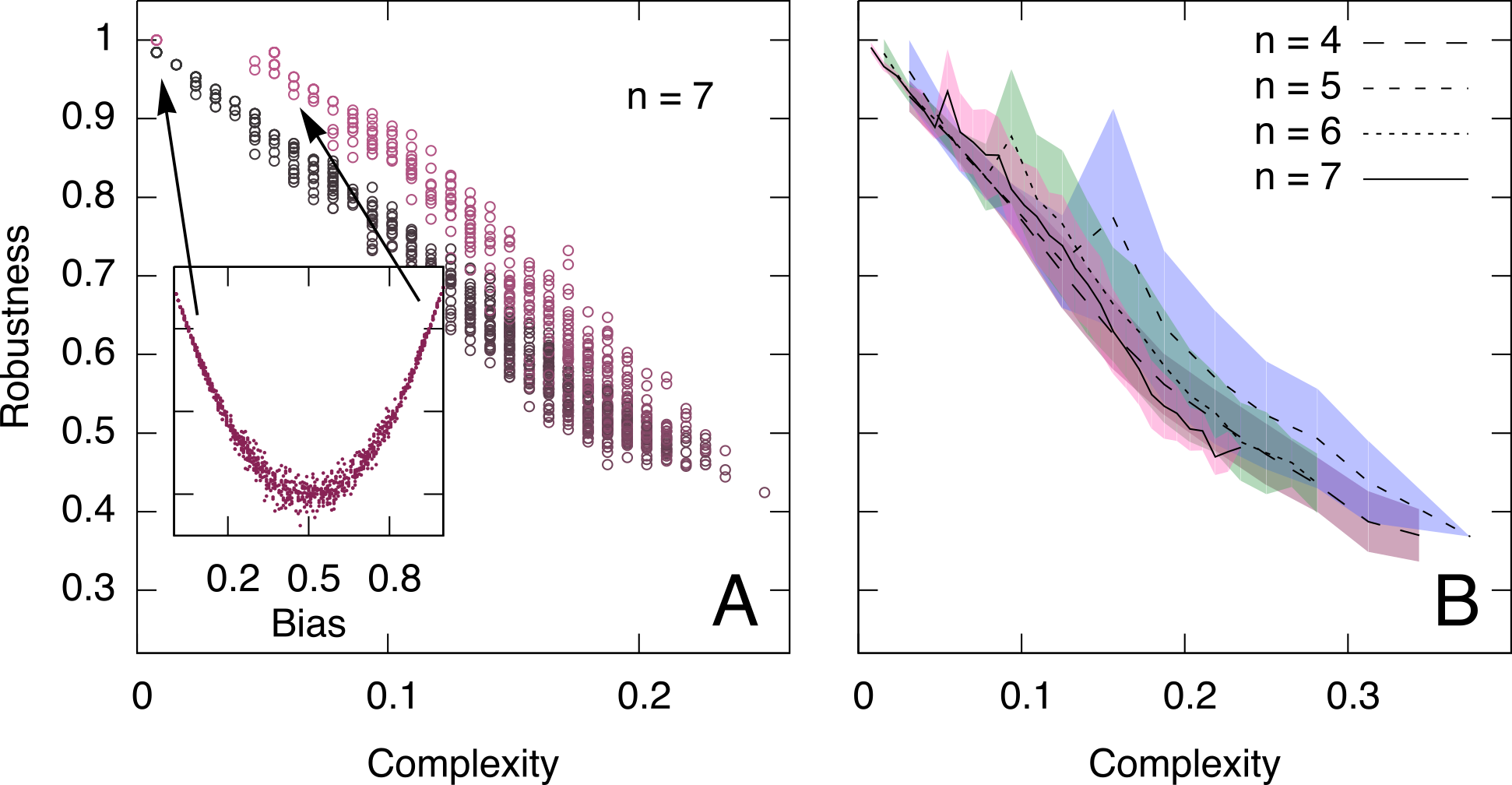}
\caption{\label{figure:robustness}
Complexity of a Boolean function ($x$ axis) strongly constrains 
its robustness ($y$ axis), while bias and number of variables have a weak influence.
Points in panel A show the robustness versus the complexity
of 1000 random Boolean functions with $n=7$ literals,
generated uniformly on the full range of biases;
the inset shows robustness versus bias for the same data.
Panel B collects data for different numbers of literals,
showing the average (lines) and the $80\%$ variability interval (shaded areas)
of the robustness as functions of the complexity.
}
\end{figure}

Noise is an important element of both living and artificial systems.
Robustness 
against errors and fluctuations,
for instance in protein folding or signal transduction,
is a central question in biology
\cite{Bialek:BOOK,Kitano:2004}.
In the field of regulatory networks, noise can be implemented by means
of a stochastic generalisation, called \emph{probabilistic} Boolean networks
\cite{ShmulevichDougherty:2002},
where the functions computed by nodes are subject to
a certain degree of variability.
One can then ask what noise level the system can sustain
without disrupting its functions.
Such questions are relevant in technological systems as well.

It is then interesting to discover that Boolean complexity
is closely related to fault tolerance
in our simple setting.
We employ the definition of robustness 
$R$ detailed in the Appendix,
which counts the fraction of single-variable flips that have no effect on the output.
More precisely, it is equal to the probability that
$f(p_1,\ldots,p_j,\ldots,p_n)=f(p_1,\ldots,\lnot p_j,\ldots,p_n)$,
when the values of $p_1,\ldots,p_n$ are chosen randomly
and $j$ is a random integer between 1 and $n$.
As discussed above, simple functions are
intuitively recognised as those having a large number of symmetries,
where
a symmetry (as outlined by the technical notion of \emph{cell}
described in the Appendix)
is a group of input combinations such that the function's value is not sensitive
to changes of some of the variables.
Thus, one expects simpler functions to be more robust.
In fact, computing the robustness for functions with varying biases
and complexities shows that $R$ is strongly dependent on $C$
(and very slightly on $B$).
Figure \ref{figure:robustness} displays these correlations,
and shows that the dependence of $R$ on the number of variables $n$
is almost undetectable.
Statistically significant correlations remain also if one
conditions the analysis to fixed values of the bias,
thus confirming the relation.
Also, flipping $2$ variables instead of $1$,
thus increasing the noise level in the definition of $R$,
has negligible effects on the results.

\section{Application to empirical systems\label{section:application}}

We are going to apply the concepts developed in the previous sections
to empirical systems of different types, belonging to the broad areas of
biology, technology and mathematics.
In particular, we will focus on transcription regulation in Eukaryotes,
digital electronic circuitry in a general-purpose processor,
and theorems in propositional logic.
The three data sets employed here are chosen as a reference,
and do not intend to be general representatives of their respective fields.
However, interesting features about
how logic complexity is expressed in empirical systems
can be isolated already from this limited exploration.

\subsection{Data sets}

\paragraph{Genetic regulation}

Transcription regulation is the machinery by which a cell coordinates
the generation of RNA from DNA, ultimately orchestrating the production
of proteins in response to internal and external stimuli.
Several proteins, including \emph{transcription factors}
that bind to the DNA, can participate in the regulation of a single gene.
They can play the simple roles of activators and repressors of transcription, 
but their complex interactions within chromatin can generate
complicated dependencies between their presence/absence patterns
and the expression level of a given gene.
These relationships can then be summarised by Boolean functions
expressing whether each gene is transcribed or not, depending
on the presence of each protein that has an influence on the gene.
We compiled a small data set of 34 such functions,
obtained from the literature
(5 regulating flower morphogenesis in \emph{Arabidopsis thaliana}
\cite{AlvaroAldana:2006};
15 regarding segment polarity in \emph{Drosophila melanogaster}
\cite{AlbertOthmer:2003};
6 controlling the mammalian cell cycle 
\cite{FaureNaldi:2006};
8 belonging to T lymphocytes in vertebrates
\cite{MendozaXenarios:2006,KlamtSaezRodriguez:2006}).
We restricted to papers where experimentally-validated functions
were employed for the construction of Boolean-network
representations of gene interactions,
since these are the most easily accessible, and they have the additional
benefit of being used already in a Boolean setting.
We circumscribe the analysis to functions with $n=3,4,5,6,7$ inputs.

\paragraph{Digital circuits}

Logic functions are the fundamental building blocks of digital electronics.
As mentioned above, hardware engineering needs were
the driving force behind the deployment of the known algorithms
for minimising Boolean functions.
Digital circuits are natively composed of logic gates,
and therefore have a natural network representation
which has been already investigated 
within a statistical physics viewpoint
\cite{CanchoJanssen:2001}.
We used data from the ITC'99 benchmark
\cite{CornoReorda:2000},
considering a partial logic-gate representation of the
Intel 80386 processor
(data set b15 \cite{ITC99}).
The data are in the form of a graph where nodes are gates
computing simple functions (either AND, NAND, OR, NOR, or NOT) of a small number of inputs,
and links run from outputs to inputs of nodes.
The full network has around 8000 nodes and 17000 links.
We built individual functions by considering all sub-graphs
with $n=3,4,5,6,7$ input links.
The enumeration was restricted to sub-graphs with
at most 5 hierarchical levels (i.e., the longest path
from an input node to an output node travels along 4 links);
the data were then pruned of functions
corresponding to a single node.
Our final data set comprises 1891 Boolean functions
(approximately 250 for $n=3$, 170 for $n=4$, and 500 for $n=5$, $n=6$, and $n=7$).

\paragraph{Formal logic}

The relations between logic functions and expressions
constitute the branch of logic known as propositional calculus.
\emph{Deductive systems} can be used to formalise and check,
solely from syntactic grounds, whether a given formula
is a consequence of another.
Basically, 
they rely on a set of \emph{axioms},
that are true by definition, and a set of \emph{inference rules},
that are used to form true expressions starting from true premises.
The data set we use is based on the Metamath project
\cite{Metamath}.
which implements a standard deductive system for the
formalisation of mathematics, providing a language and
a proof-validation software to the community of people involved.
We restricted to the part of the Metamath database that regards
propositional calculus, for which one can interpret expressions
as Boolean functions.
It depends on only three axioms,
(known as the principles of \emph{simplification}, \emph{transposition},
and \emph{Frege}) and only one inference rule, the \emph{modus ponens}.
Theorems can be in either one of two forms.
The first is $\vdash P$ --- where $P$ is an expression --- meaning
``$P$ is provable in the formal system'', in which case
$P$ is a tautology, thanks to the coherence of the system.
The second is $\vdash Q_1, \vdash Q_2, \ldots, \vdash Q_k \Rightarrow \vdash R$,
meaning ``if all $Q_i$'s are provable in the system, than so is $R$''.
Our data set was constructed as follows.
If a theorem is in the second class, we keep the proposition $R$.
If it is in the first class, it is trivially a tautology (maximum bias, minimum complexity),
so we parse $P$ and cast it into
the form $OP(Q,R)$, where OP is a binary operator;
then if OP is a conditional ($Q\rightarrow R$) we keep $R$
(since the original theorem could have been written
as $\vdash Q \Rightarrow \vdash R$),
if OP is a biconditional ($Q\leftrightarrow R$) we keep both $Q$ and $R$,
if OP is a disjunction ($Q\lor R$) we keep both $Q$ and $R$
(since it could have been written as
$\vdash \lnot Q \Rightarrow \vdash R$
and $\vdash \lnot R \Rightarrow \vdash Q$).
We end up with 327 propositions.

\subsection{Empirical functions have low complexity}

\begin{figure}
\centering
\includegraphics[scale=1.1]{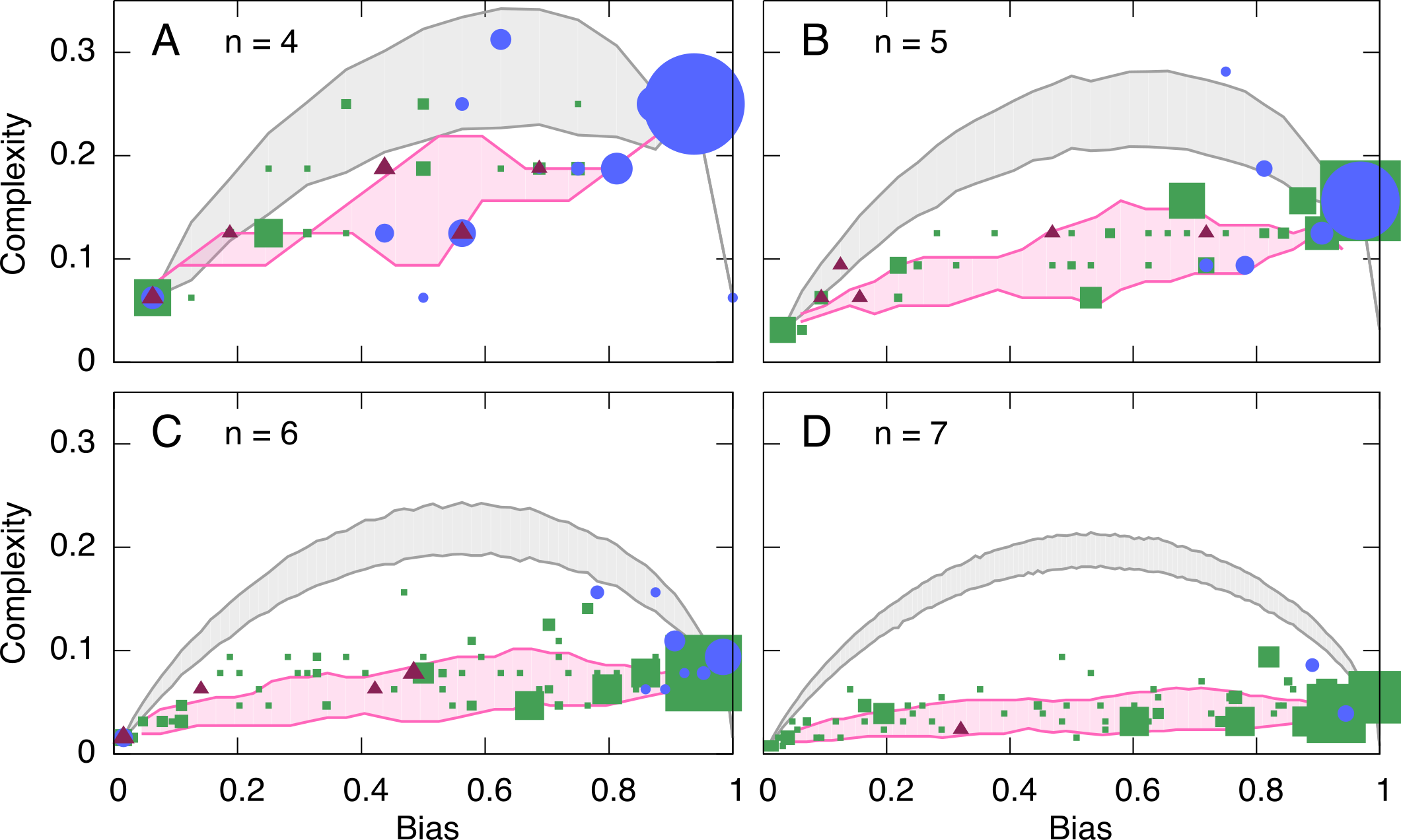}
\raisebox{0.23\height}{
\includegraphics[scale=0.47]{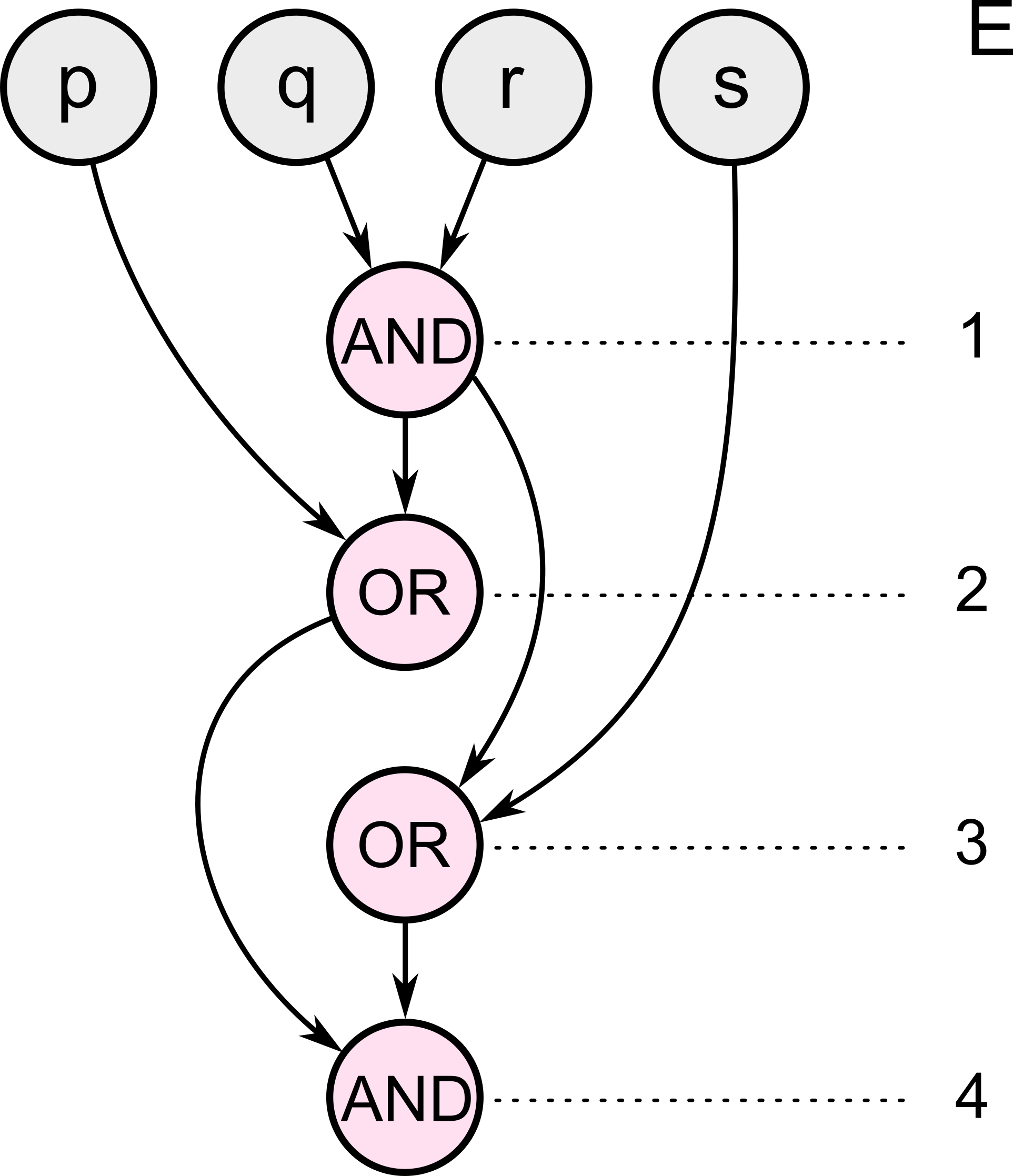}}
\caption{\label{figure:morphospace}
The bias (on $x$ axes) 
and the complexity (on $y$ axes)
of logic functions from empirical data sets
are related in a non-null way, which is captured
by a model of random Boolean networks.
(A,B,C,D)
Functions defined by 
genetic regulatory circuits (red triangles), 
electronic sub-circuits in the i386 microprocessor (green squares),
and theorems in propositional logic (blue circles)
have lower complexity than typical Boolean functions,
especially for larger numbers of literals.
Grey shaded areas comprise $\sim80\%$ of all functions with a given bias.
Point sizes are proportional to the number of data points
having the same bias and complexity.
The pink shaded areas are small random Boolean networks,
schematically described in panel (E).
In this example, the network is built in four steps;
the resulting function is given by the bottommost node
(which is the first one that connects to all literals $p,q,r,s$),
and is $(p\lor(p\land q))\land(s\lor(p\land q))$.
}
\end{figure}

The values of bias and complexity for the function
in our data sets are shown in Fig.~\ref{figure:morphospace}A--D.
The bias (on the horizontal axis)
discriminates between the three systems, for each $n$ considered.
Functions expressing gene-regulatory rules take the value $0$
(meaning ``no gene transcript'') more often than in the other systems, 
while theorems in mathematical logic show an inclination for the value $1$
(meaning ``true'');
electronic sub-circuits, though more balanced,
are slightly biased towards the ``on'' state,
contrary to what one would expect from energy-consumption considerations.
Comparison with the null model shows that the complexity of empirical functions
is consistently lower than the typical Boolean functions,
especially for larger $n$.

\subsection{A random-network model reproduces the empirical trends}

A noticeable feature of the empirical B-C plots is the monotonic
correlation between bias and complexity, in spite of the non-monotonic
one expected for random functions.
This suggests the existence of a positive mechanism
present either in the empirical systems themselves
or in the way they are modelised through logic functions.
We explore a possible scenario based on a model of
\emph{random Boolean networks}, defined below.
Our goal is to show that analysis of empirical data
by means of the bias-complexity coordinates
can be helpful in discerning non-null features in the data
and devising positive models.

Let us start with $n$ input nodes, which represent the literals
(refer to Fig.~\ref{figure:morphospace}E).
We build a graph iteratively by adding nodes one by one.
Each new node carries a random binary function $\phi$
(either OR or AND),
and attaches itself to two randomly chosen nodes,
among those already present (literals included).
If the values of these nodes are $p$ and $q$,
than the value of the new node will be $\phi(p,q)$.
The growth process is stopped as soon as
the first node depending on all $n$ input literals appears.
More precisely, the function $f$ embodied by the network
is defined as the first node whose \emph{light cone}
contains all input nodes,
the light cone of a node $A$ being defined as the set of all
nodes $B$ such that there is a directed path in the network
going from $A$ to $B$.
Figure~\ref{figure:morphospace} shows that the
bias-complexity relation predicted by this model
is in reasonable accord with the trend observed
in the empirical systems.
The most notable feature is the roughly linear
correlation between the two coordinates,
which is instead non-monotonic for random functions.

Notice that we have chosen $\phi$ to be either OR or AND
with equal probability.
This choice generates functions with biases covering
the whole interval $(0,1)$; a statistical
prevalence of OR reduces the average $B$,
while the opposite happens for AND,
without modifying the trend observed.
Adding a negation (NOT) in front of randomly-chosen
variables does not change the results appreciably.

\section{Discussion}

The view presented here is based on the observation that
logic functions are a widespread tool in modelling complex systems,
realising compact two-state descriptions of complicated response functions.
%
Therefore, a definition of complexity in a Boolean setting
is useful, as it enables the quantitative comparison of behaviours
across systems. 
It proves fruitful also within fields,
as a concise indicator summarising several 
``microscopic'' features in a single global observable.

As we showed, complexity effectively discriminates different classes of functions
widely used in modelling approaches.
Specifically within the class of canalising functions, 
it produces a quantitative measure of ``how much'' canalisation is realised.
This measure is more general and more detailed than the number
of canalising variables; it can be measured exactly for all Boolean functions
and is arguably more suited to information-theoretic analyses.
A long-standing question in the field of genetic regulation concerns the properties
of regulatory networks responsible for their not being chaotic
\cite{Kauffman:BOOK,AldanaCoppersmith:2003}.
It is well known
that Boolean networks can
display chaotic dynamics in certain regimes, at variance with the ordered state
their are found to be in living systems.
Canalisation is
one quality of regulatory rules that has been found to promote
network stability.
It would be interesting to investigate how the
order-chaos transition in the Kauffman model
(a random network of random Boolean functions
\cite{Kauffman:1969})
depends on the complexity of the rules;
the results regarding robustness described above are
relevant in this sense
(see also 
\cite{MarquesPitaRocha:2011,KauffmanPeterson:2004,KauffmanPeterson:2003}).
Another consequence of our results on the classes
of functions commonly used for gene regulation
is that threshold functions
impose a systematic tendency
towards high complexity, and constrain both
bias and complexity to very specific values.

The relations uncovered between complexity, 
on/off asymmetry, and robustness 
seem to indicate the presence of
architectural similarities between the empirical systems considered.
In particular, the low Boolean complexity observed
in our three data sets and the monotonic correlation between $C$ and $B$
suggest the existence of a positive mechanism
underlying these empirical systems.
A possible rationalisation is 
given by the simplified model of random logic networks
described above,
which reproduces the trends.
The information provided is twofold.
First, our example of an empirical application
shows that measuring complexity,
especially combined with bias in the B-C parametrisation,
can promote the discovery of hidden regularities, trends, and tradeoffs.
Second, it suggests a possible underlying mechanism
recapitulating the statistical regularities measured,
namely a modular structure where the whole function
to be realised by the system is expressed by means of smaller functions
(i.e., with fewer inputs than $n$) organised
in a hierarchical arrangement.
Such a common architecture need not be generated
by common evolutionary processes in the three systems.
It may be the consequence of selection ---
for instance favouring robustness ---,
or a neutral
effect of the system's organisation, or it could expose
our preference for simple structures,
at least in artificial systems.
Remarkably, Boolean complexity of concepts has been found to be
a predictor of subjective difficulty in human learning
\cite{Feldman:2000}.
(In the case of regulation, the difficulty in performing the
experiments and completely identifying the set of regulating
proteins may be partially responsible for the low complexities observed.)

Finally, we remark that the results presented
are not sensitive on the particular description language
one employs in the definition of logic complexity.
We used
\emph{disjunctive} normal forms (i.e., disjunctions of conjunctive clauses)
throughout the paper, but we checked that definitions based on
\emph{conjunctive} normal forms (conjunctions of disjunctive clauses) or
\emph{algebraic} normal forms (exclusive disjunctions of conjunctive clauses)
do not change our result statements.

\section*{Acknowledgements}
We are thankful to Marco Cosentino Lagomarsino
for discussions and useful comments on a previous version of the manuscript.
We also acknowledge discussions with
Bruno Bassetti and Jacopo Grilli.


\appendix

\section*{Appendix}

A \emph{Boolean function} $f$ of $n$ variables is a map
\begin{equation*}
\begin{split}
f:\{0,1\}^n&\to \{0,1\}\\
\sigma&\mapsto f(\sigma)
\end{split}
\end{equation*}
where $\sigma=\{p_1,p_2,\ldots,p_n\}$ is the configuration
of the input variables (or \emph{literals}) $p_i$.
A logic function $f(p_1,p_2,\ldots,p_n)$ is uniquely determined
by its \emph{truth table}, which is the explicit listing of the value of $f$
for all $2^n$ possible combinations of the $n$ variables $p_i$.
Notable functions are the \emph{tautology},
which takes the value 1 for all $\sigma$
(a tautology is ``always true''),
and the \emph{parity function}
$\pi(p_1,\ldots,p_n)=\sum p_i \;\mathrm{mod}\; 2$,
which takes the value 1 when the number of literals
equal to 1 is odd (and 0 otherwise).

Boolean functions can be built by composition of
``smaller'' functions, e.g.\ a ternary function $f$
can be obtained starting from two binary functions
$g$ and $h$ as $f(p,q,r):=g(p,h(q,r))$.
This permits the definition of a small ``vocabulary''
of atomic functions, from which
\emph{Boolean expressions}, corresponding to
larger functions, can be constructed by composition.
As the basic building blocks for logic functions, we choose
the two binary functions AND ($\land$) and OR ($\lor$),
and the unary function NOT ($\lnot$).
The set $\mathcal B=\left\{\land,\lor,\lnot\right\}$ is functionally complete, 
meaning that the three atomic functions
can be composed to represent
all possible functions of $n$ literals.
Notice that $\mathcal B$ is not minimal, as both $\land$ and $\lor$
could be expressed by means of the other two connectives.
However, it is a very natural generating set, as it is tightly linked
to the representation of functions in terms of their truth table.

The two-dimensional \emph{morphospace} we use
summarises each function by two quantities:
bias and complexity.
Bias measures the average value taken by the function,
and is therefore an indicator of the asymmetry
between the ``on'' (1) and ``off'' (0) output states.

\begin{definition}
The \emph{bias} $B$ (or \emph{on/off asymmetry}) of a Boolean function $f$
is the fraction of input combinations for which $f$ is true, namely
\begin{equation*}
B=\sum_{\{\sigma\}}f(\sigma)/\sum_{\{\sigma\}}1,
\end{equation*}
where the sums are over all possible input combinations $\sigma$
(hence $\sum 1=2^n$ for a function of $n$ variables).
\end{definition}

Defining complexity, as in Kolmogorov's definition,
requires the specification of a description language.
A natural language for logic functions is that of
\emph{disjunctive normal forms} (DNF).
As described in the text, these are Boolean expressions built with the elements 
of our base set $\mathcal B$, such that they take a canonical form,
namely a disjunction of conjunctive clauses, in terms of literals and their negations.
A DNF is said to be \emph{full} if each literal appears exactly once
in each conjunctive clause.
If a logic function $f$ is expressed via its truth table, then its
full DNF can be immediately constructed,
simply by listing all combinations of truth values for which $f$ is true.
For instance, the function XOR$(p,q)$,
which is true if exactly one of its literals $p,q$ is true,
can be written as $(p\land\lnot q)\lor(\lnot p\land q)$.
Let us define
the \emph{length} $l(\varphi)$ of a DNF $\varphi$ as the number of clauses,
i.e., the number of terms separated by $\lor$ operators).
The bias of the function is then equal to
the length of the full DNF.
Each term in a DNF can be considered as an autonomous sub-function
of a number $k$ of literals,
defining a \emph{k-cell} in the function's truth table.
A $k$-cell of a function is a subset of input combinations
for which the function is true and such that it can be expressed,
when restricted to those combinations,
as a single conjunctive term $p_1\land p_2\land\ldots\land p_k$
(possibly with negations),
all other $n-k$ literals remaining free.
The full disjunctive normal form is typically not a compact
representation of a function.
As an extreme case, 
consider the \emph{tautology} of $n$ variables:
its bias is $1$ and its full DNF has $2^n$ terms.
However, a much shorter DNF is for instance $p\lor\lnot p$
(where $p$ is any literal), which is still in DNF and has length 2,
thus defining the same function with a much shorter expression.
As discussed in the text, complexity is expected to measure
the minimal amount of information one has to specify when defining
an entity.
Having fixed DNFs as the natural language,
it is then straightforward to define the complexity of a logic function
as follows.

\begin{definition}
Let us denote by $\Phi(f)$ the set of all the DNFs of the function $f$
(it is a finite set if repetitions of clauses are prohibited).
The \emph{complexity} $C$ of a Boolean function is the length
of its shortest disjunctive normal form, normalised by the length of
its truth table, namely
\begin{equation*}
C=\min_{\varphi\in\Phi(f)} l(\varphi)/\sum_{\{\sigma\}} 1,
\end{equation*}
where $\sigma$ is as in Definition~1.
\end{definition}

An equivalent definition in terms of cells can be given,
as the minimum number of cells needed to cover the
set of input combinations for which the function is 1
(see e.g.\ \cite{Carbone:BOOK}).

Since the full DNF belongs to $\Phi(f)$, and its length 
is the number of 1s in the truth table,
the inequality $C\leq B$ holds in general.
It is interesting to note that the parity function of $n$ variables,
for which $B=1/2$,
has the largest possible complexity, namely $C=1/2$
(the sequence of its values realises a fractal
known as the \emph{Thue-Morse sequence}).
This is easily proved by noting that if there existed any cell
containing more than one element, than it would contain
at least two combinations of inputs having different parities.
Parity functions are used in various contexts, due to their
symmetry and tractability \cite{CiandriniMaffi:2009,MezardMontanari:BOOK}.

Finally, the notion of robustness measures how much fluctuations
in the input variables affect the function's value.
\begin{definition}
The \emph{robustness} $R$ of a Boolean function $f$ is the fraction of
pairs $\{\sigma_1,\sigma_2\}$ such that $f(\sigma_1) = f(\sigma_2)$,
where $\sigma_1$ and $\sigma_2$
are two combinations of inputs differing only in the value of one variable
(i.e., their Hamming distance $|\sigma_1-\sigma_2|$ is 1), namely
\begin{equation*}
R=\sum_{|\sigma_1-\sigma_2|=1}\!\!
\delta_{f(\sigma_1),f(\sigma_2)}
/\!\!\sum_{|\sigma_1-\sigma_2|=1} \!\!\!1
\end{equation*}
\end{definition}
Tautologies and their negations have the highest robustness, namely $R=1$,
as changing the value of any variable never changes the result.
The parity function, on the contrary, has the lowest robustness, $R=0$,
since by definition a single flip of any of its variables changes the
function's value.
Remark that this definition of $R$ fixes a specific
scale for the fluctuations, namely only $1$ variable.
The analogous definition where one considers $2$ flips would
assign minimum complexity to parity functions.
However, the results presented in the text are unaffected
by the number of variables flipped, showing that the relation
between robustness and complexity is robust.



\begin{thebibliography}{10}
\expandafter\ifx\csname url\endcsname\relax
  \def\url#1{\texttt{#1}}\fi
\expandafter\ifx\csname urlprefix\endcsname\relax\def\urlprefix{URL }\fi
\expandafter\ifx\csname href\endcsname\relax
  \def\href#1#2{#2} \def\path#1{#1}\fi

\bibitem{Carbone:BOOK}
A.~Carbone, S.~Semmes, A Graphic Apology for Symmetry and Implicitness, Oxford
  mathematical monographs, Oxford University Press, 2000.

\bibitem{SteinNewman:BOOK}
D.~Stein, C.~Newman, Spin Glasses and Complexity, Primers in Complex Systems,
  Princeton University Press, 2013.

\bibitem{BarratBarthelemyVespignani:BOOK}
A.~Barrat, M.~Barthlemy, A.~Vespignani, Dynamical Processes on Complex
  Networks, 1st Edition, Cambridge University Press, New York, NY, USA, 2008.

\bibitem{HidalgoHausmann:2009}
C.~A. Hidalgo, R.~Hausmann, The building blocks of economic complexity,
  Proceedings of the National Academy of Sciences 106~(26) (2009) 10570--10575.

\bibitem{TacchellaCristelli:2012}
A.~Tacchella, M.~Cristelli, G.~Caldarelli, A.~Gabrielli, L.~Pietronero,
  \href{http://dx.doi.org/10.1038/srep00723}{A new metrics for countries'
  fitness and products' complexity}, Sci. Rep. 2.

\bibitem{TononiEdelman:1998}
G.~Tononi, G.~M. Edelman, Consciousness and complexity, Science 282~(5395)
  (1998) 1846--1851.

\bibitem{McNerneyFarmer:2011}
J.~McNerney, J.~D. Farmer, S.~Redner, J.~E. Trancik,
  \href{http://www.pnas.org/content/108/22/9008.abstract}{Role of design
  complexity in technology improvement}, Proceedings of the National Academy of
  Sciences 108~(22) (2011) 9008--9013.

\bibitem{Frenken:2006}
K.~Frenken, Technological innovation and complexity theory, Economics of
  Innovation and New Technology 15~(2) (2006) 137--155.

\bibitem{FeldmanCrutchfield:1998}
D.~P. Feldman, J.~P. Crutchfield, Measures of statistical complexity: Why?,
  Physics Letters A 238~(4--5) (1998) 244 -- 252.

\bibitem{ShaliziShalizi:2004}
C.~R. Shalizi, K.~L. Shalizi, R.~Haslinger,
  \href{http://link.aps.org/doi/10.1103/PhysRevLett.93.118701}{Quantifying
  self-organization with optimal predictors}, Phys. Rev. Lett. 93 (2004)
  118701.

\bibitem{AuerbachBongard:2014}
J.~E. Auerbach, J.~C. Bongard, Environmental influence on the evolution of
  morphological complexity in machines, PLoS Comput Biol 10~(1) (2014)
  e1003399.

\bibitem{JoshiTononi:2013}
N.~J. Joshi, G.~Tononi, C.~Koch,
  \href{http://dx.doi.org/10.1371\%2Fjournal.pcbi.1003111}{The minimal
  complexity of adapting agents increases with fitness}, PLoS Comput Biol 9~(7)
  (2013) e1003111.

\bibitem{WangLiao:2010}
Z.~Wang, B.-Y. Liao, J.~Zhang,
  \href{http://www.pnas.org/content/107/42/18034.abstract}{Genomic patterns of
  pleiotropy and the evolution of complexity}, Proceedings of the National
  Academy of Sciences 107~(42) (2010) 18034--18039.

\bibitem{AdamiOfria:2000}
C.~Adami, C.~Ofria, T.~C. Collier,
  \href{http://www.pnas.org/content/97/9/4463.abstract}{Evolution of biological
  complexity}, Proceedings of the National Academy of Sciences 97~(9) (2000)
  4463--4468.

\bibitem{CseteDoyle:2002}
M.~E. Csete, J.~C. Doyle,
  \href{http://www.sciencemag.org/content/295/5560/1664.abstract}{Reverse
  engineering of biological complexity}, Science 295~(5560) (2002) 1664--1669.

\bibitem{CarlsonDoyle:2002}
J.~Carlson, J.~Doyle, Complexity and robustness, Proceedings of the National
  Academy of Sciences 99~(suppl 1) (2002) 2538--2545.

\bibitem{AlbertJeong:2000}
R.~Albert, H.~Jeong, A.-L. Barabasi,
  \href{http://dx.doi.org/10.1038/35019019}{Error and attack tolerance of
  complex networks}, Nature 406~(6794) (2000) 378--382.

\bibitem{MaciaSole:2009}
J.~Macia, R.~V. Sol{\'e}, Distributed robustness in cellular networks: insights
  from synthetic evolved circuits, Journal of The Royal Society Interface
  6~(33) (2009) 393--400.

\bibitem{LiLong:2004}
F.~Li, T.~Long, Y.~Lu, Q.~Ouyang, C.~Tang,
  \href{http://www.pnas.org/content/101/14/4781.abstract}{The yeast cell-cycle
  network is robustly designed}, Proceedings of the National Academy of
  Sciences of the United States of America 101~(14) (2004) 4781--4786.

\bibitem{Wegener:BOOK}
I.~Wegener, The Complexity of Boolean Functions, John Wiley \&amp; Sons, Inc.,
  New York, NY, USA, 1987.

\bibitem{TkacikBialek:2014}
G.~Tka{\v c}ik, W.~Bialek, \href{http://arxiv.org/abs/1412.8752}{Information
  processing in biological systems},
  Annu Rev Cond Matt Phys 7 (2016)

\bibitem{BarveWagner:2013}
A.~Barve, A.~Wagner, \href{http://dx.doi.org/10.1038/nature12301}{A latent
  capacity for evolutionary innovation through exaptation in metabolic
  systems}, Nature 500~(7461) (2013) 203--206.

\bibitem{SoleValverde:2013}
R.~V. Sol{\'e}, S.~Valverde, M.~R. Casals, S.~A. Kauffman, D.~Farmer,
  N.~Eldredge, \href{http://dx.doi.org/10.1002/cplx.21436}{The evolutionary
  ecology of technological innovations}, Complexity 18~(4) (2013) 15--27.

\bibitem{HarrisSawhill:2002}
S.~E. Harris, B.~K. Sawhill, A.~Wuensche, S.~Kauffman,
  \href{http://dx.doi.org/10.1002/cplx.10022}{A model of transcriptional
  regulatory networks based on biases in the observed regulation rules},
  Complexity 7~(4) (2002) 23--40.

\bibitem{Bowers:2004}
P.M.~Bowers, S.J.~Cokus, D.~Eisenberg, T.O.~Yeates,
{Use of logic relationships to decipher protein network organization},
Science 306 (2004) 2246--9.


\bibitem{Carlet:2010}
C.~Carlet, \href{http://dx.doi.org/10.1017/CBO9780511780448.011}{Boolean
  functions for cryptography and error-correcting codes}, in: Y.~Crama, P.~L.
  Hammer (Eds.), Boolean Models and Methods in Mathematics, Computer Science,
  and Engineering, Cambridge University Press, 2010, pp. 257--397, cambridge
  Books Online.

\bibitem{Micheli:BOOK}
G.~D. Micheli, Synthesis and Optimization of Digital Circuits, 1st Edition,
  McGraw-Hill Higher Education, 1994.

\bibitem{Bilbao:BOOK}
J.~Bilbao, \href{http://books.google.es/books?id=AN3JqmxaC-gC}{Cooperative
  Games on Combinatorial Structures}, Theory and Decision Library C, Springer
  US, 2000.

\bibitem{Church:BOOK}
A.~Church, \href{http://opac.inria.fr/record=b1107343}{Introduction to
  mathematical logic. vol. 1} (1956).

\bibitem{GongSocolar:2012}
X.~Gong, J.~E.~S. Socolar,
  \href{http://link.aps.org/doi/10.1103/PhysRevE.85.066107}{Quantifying the
  complexity of random boolean networks}, Phys. Rev. E 85 (2012) 066107.

\bibitem{CiandriniMaffi:2009}
L.~Ciandrini, C.~Maffi, A.~Motta, B.~Bassetti, M.~Cosentino Lagomarsino,
  \href{http://arxiv.org/abs/0809.1526}{Feedback topology and {XOR}-dynamics in
  {B}oolean networks with varying input structure}, Phys Rev E 80 (2009)
  026122.

\bibitem{MarquesPitaMitchell:2008}
M.~Marques-Pita, M.~Mitchell, L.~Rocha,
  \href{http://dx.doi.org/10.1007/978-3-540-85194-3_13}{The role of conceptual
  structure in designing cellular automata to perform collective computation},
  in: 
  Proceedings of the 7th international conference on unconventional computing, 
  2008, pp. 146--163.

\bibitem{MarquesPitaRocha:2011}
M.~Marques-Pita, L.~Rocha, Schema redescription in cellular automata:
  Revisiting emergence in complex systems, in: Artificial Life (ALIFE), 2011
  IEEE Symposium on, 2011, pp. 233--240.

\bibitem{Nakamura:1979}
K.~Nakamura, The vetoers in a simple game with ordinal preferences,
  International Journal of Game Theory 8 (1979) 55--61.

\bibitem{Papadimitriou:2014}
C.~Papadimitriou,
  \href{http://www.pnas.org/content/111/45/15881.abstract}{Algorithms,
  complexity, and the sciences}, Proceedings of the National Academy of
  Sciences 111~(45) (2014) 15881--15887.

\bibitem{Korte:BOOK}
B.~Korte, J.~Vygen, Combinatorial Optimization: Theory and Algorithms, Springer
  Publishing Company, Incorporated, 2007.

\bibitem{Espresso}
R.~Rudell, A.~Sangiovanni-Vincentelli, Multiple-valued minimization for pla
  optimization, Computer-Aided Design of Integrated Circuits and Systems, IEEE
  Transactions on 6 (1987) 727--750.

\bibitem{Kauffman:BOOK}
S.~A. Kauffman, The Origins of Order: Self-Organization and Selection in
  Evolution, 1st Edition, Oxford University Press, USA, 1993.

\bibitem{Davidich:2008}
M.~Davidich, S.~Bornholdt, Boolean network model predicts cell cycle sequence
  of fission yeast, PLoS ONE 3~(2) (2008) e1672.

\bibitem{KauffmanPeterson:2003}
S.~Kauffman, C.~Peterson, B.~Samuelsson, C.~Troein,
  \href{http://www.pnas.org/content/100/25/14796.abstract}{Random boolean
  network models and the yeast transcriptional network}, Proceedings of the
  National Academy of Sciences 100~(25) (2003) 14796--14799.

\bibitem{Bialek:BOOK}
W.~Bialek, Biophysics: Searching for Principles, Princeton University Press,
  2012.

\bibitem{Kitano:2004}
H.~Kitano, \href{http://dx.doi.org/10.1038/nrg1471}{Biological robustness}, Nat
  Rev Genet 5~(11) (2004) 826--837.

\bibitem{ShmulevichDougherty:2002}
I.~Shmulevich, E.~R. Dougherty, S.~Kim, W.~Zhang,
  \href{http://bioinformatics.oxfordjournals.org/content/18/2/261.abstract}{Probabilistic
  {B}oolean networks: a rule-based uncertainty model for gene regulatory
  networks}, Bioinformatics 18~(2) (2002) 261--274.

\bibitem{AlvaroAldana:2006}
{\'A}.~Chaos, M.~Aldana, C.~Espinosa-Soto, B.~de~Le{\'o}n, A.~Arroyo,
  E.~Alvarez-Buylla, \href{http://dx.doi.org/10.1007/s00344-006-0068-8}{From
  genes to flower patterns and evolution: Dynamic models of gene regulatory
  networks}, Journal of Plant Growth Regulation 25~(4) (2006) 278--289.

\bibitem{AlbertOthmer:2003}
R.~Albert, H.~G. Othmer, The topology of the regulatory interactions predicts
  the expression pattern of the segment polarity genes in drosophila
  melanogaster, J Theor Biol 223 (2003) 1--18.

\bibitem{FaureNaldi:2006}
A.~Faure, A.~Naldi, C.~Chaouiya, D.~Thieffry, Dynamical analysis of a generic
  {B}oolean model for the control of the mammalian cell cycle, Bioinformatics
  22~(14) (2006) e124--31+.

\bibitem{MendozaXenarios:2006}
L.~Mendoza, I.~Xenarios, \href{http://dx.doi.org/10.1186/1742-4682-3-13}{A
  method for the generation of standardized qualitative dynamical systems of
  regulatory networks}, Theoretical Biology and Medical Modelling 3~(1).

\bibitem{KlamtSaezRodriguez:2006}
S.~Klamt, J.~Saez-Rodriguez, J.~Lindquist, L.~Simeoni, E.~D. Gilles, {A
  methodology for the structural and functional analysis of signaling and
  regulatory networks}, BMC Bioinformatics 7 (2006) 56.

\bibitem{CanchoJanssen:2001}
R.~Ferrer i Cancho, C.~Janssen, R.~V. Sol\'e,
  \href{http://link.aps.org/doi/10.1103/PhysRevE.64.046119}{Topology of
  technology graphs: Small world patterns in electronic circuits}, Phys. Rev. E
  64 (2001) 046119.

\bibitem{CornoReorda:2000}
F.~Corno, M.~Reorda, G.~Squillero, {RT-level ITC'99 benchmarks and first ATPG
  results}, Design Test of Computers, IEEE 17~(3) (2000) 44--53.

\bibitem{ITC99}
\href{http://www.cad.polito.it/downloads/tools/itc99.html}{Itc'99 benchmarks
  (2nd release)} [online, cited 5 Jan 2015].
  \newline\urlprefix\url{http://www.cad.polito.it/downloads/tools/itc99.html}

\bibitem{Metamath}
\href{http://us.metamath.org}{Metamath home page} [online, cited 6 May 2015].
  \newline\urlprefix\url{http://us.metamath.org}

\bibitem{AldanaCoppersmith:2003}
M.~Aldana, S.~Coppersmith, L.~P. Kadanoff, Boolean dynamics with random
  couplings, Springer-Verlag, 2003, pp. 23--89.

\bibitem{Kauffman:1969}
S.~A. Kauffman, Metabolic stability and epigenesis in randomly constructed
  genetic nets, Journal of theoretical biology 22~(3) (1969) 437--467.

\bibitem{KauffmanPeterson:2004}
S.~Kauffman, C.~Peterson, B.~Samuelsson, C.~Troein,
  \href{http://www.pnas.org/content/101/49/17102.abstract}{Genetic networks
  with canalyzing boolean rules are always stable}, Proceedings of the National
  Academy of Sciences of the United States of America 101~(49) (2004)
  17102--17107.

\bibitem{Feldman:2000}
J.~Feldman, Minimization of {B}oolean complexity in human concept learning, Nature
  407 (2000) 630--632.

\bibitem{MezardMontanari:BOOK}
M.~Mezard, A.~Montanari, Information, Physics, and Computation, Oxford
  University Press, Inc., New York, NY, USA., 2009.

\end{thebibliography}
\end{document}